\documentclass[conference,10pt]{IEEEtran}
\IEEEoverridecommandlockouts

\usepackage{textcomp}
\usepackage{gensymb}
\usepackage{multirow}
\usepackage[ruled,linesnumbered,noend]{algorithm2e} 

\usepackage{cite}
\usepackage{subcaption}
\usepackage{amsmath,amssymb,amsfonts}
\usepackage[ruled,linesnumbered,noend]{algorithm2e} 
\usepackage{graphicx}
\usepackage{textcomp}

\usepackage{dblfloatfix}

\usepackage{tabularx}

\usepackage{url}
\def\BibTeX{{\rm B\kern-.05em{\sc i\kern-.025em b}\kern-.08em
    T\kern-.1667em\lower.7ex\hbox{E}\kern-.125emX}}

\begin{document}

\title{Reliable THz Communications for Outdoor based Applications- Use Cases and Methods \\
\thanks{This project was funded by CMU Portugal Program: CMU/TMP/0013/2017- THz Communication for Beyond 5G Ultra-fast Networks.}
}

\author{\IEEEauthorblockN{Rohit Singh \textsuperscript{1}, Douglas Sicker \textsuperscript{1,2}}
\IEEEauthorblockA{ \textsuperscript{1}  \textit{Engineering \& Public Policy}, \textit{Carnegie Mellon University}, Pittsburgh, USA \\
			       \textsuperscript{2}  \textit{School of Computer Science}, \textit{Carnegie Mellon University}, Pittsburgh, USA \\
Email: rohits1@andrew.cmu.edu, sicker@cmu.edu}
}

\maketitle

\section*{ABSTRACT}

Future (beyond 5G) wireless networks will demand high throughput and low latency and would benefit from greenfield, contiguous, and wider bandwidth, all of which THz spectrum can provide. Although THz has been envisioned to be deployed in an indoor setting, with proper enforcement and planning, we can draw a limited number of use cases for outdoor THz communication. THz can provide high capacity and ultra-high throughput but at the cost of high path loss and sensitivity to device orientation/mobility.. We identify scenarios where the use of the THz spectrum for an outdoor setting is justified and their critical operating parameters. We further categorize the applications based on the relative mobility between the access point (AP) and user equipment (UE). We present an approach for deploying THz on an outdoor framework by presenting preliminary technical parameter analysis for scenarios, like wireless backhaul, high-speed kiosks, and the aerial base station (ABS). Our preliminary analysis shows that the application for each of these scenarios is limited based on multiple parameters, such as distance, device mobility, device orientation, user geometry, antenna gain, and environment settings, which requires separate consideration and optimization.  

\section*{Keywords}  Outdoor mobility, Terahertz, wireless backhaul, ultra-high kiosks, aerial base station

\section{Introduction}

The 2015 International Telecommunication Union-Radiocommunication (ITU-R) report  \cite{ITUR} predicts that by the year 2030, there will be $97 billion$ connected mobile devices, which would include smartphones, tablets, and wearable. This shows that the growth of devices, and hence, the demand is superlinear. 5G spectrum, like Citizens Broadband Radio Service (CBRS) ($3.5GHz$), alone will not be able to support this increased demand for the already ultra-dense deployment of devices \cite{OurITS}. It seems likely that in the future, we will require an additional spectrum to boost the rising demand for ultra-high throughput and ultra-dense networks. The question is, where can we get easier access to vast amounts of the spectrum? Traditionally, we have used the lower frequencies, like the radio-frequency (RF) band, for outdoor mobile applications, mostly because of its higher coverage, higher penetration power, and mobility resilience. However, the RF spectrum is \textit{brown-field}, i.e., the spectrum is already allocated and requires repurposing for additional spectrum. On the other hand, the higher frequencies, like millimeter wave ($30GHz-300GHz$), THz ($100GHz-3THz$) \cite{FCCTHz} and infrared ($300GHz-430THz$) bands, the spectrum are \textit{green-field}, i.e., lightly used and requires minimal remodeling \cite{OurITS}. Higher frequencies can enable higher throughput but at the cost of higher path loss, need for beam alignment, smaller coverage area, and sensitivity to mobility, which are not idle for an outdoor type communication \cite{OurITS, Review81}. 

Although higher frequency bands offer many challenges, in this paper, we identify the characteristics and parameters required for reliable outdoor communication at higher frequencies. We propose that through scenario-wise parameter modeling, outdoor THz deployment could be possible. The THz system should adaptively change its parameters (operating frequency, available bandwidth, transmit power, antenna beamwidth, and antenna angle) based on environmental conditions (humidity, rain, wind, blockages, the distance between antennas, and UE mobility/orientation) \cite{OurGC}. We identify multiple scenarios where the use of THz spectrum is justified, like wireless backhaul, Augmented Reality (AR) based applications, quasi-mobile applications, high-speed kiosks, mobile ad-hoc networks (MANET) and Aerial Base Stations (ABS). Furthermore, we categorize these scenarios based on the relative mobility of the APs and UEs to identify the sensitive parameters. We perform a preliminary analyze for a subset of these applications: (a) wireless backhaul, (b) high-speed Kiosks, and (c) aerial base station (ABS).

\setlength\belowcaptionskip{-0.2 in}
\setlength{\abovecaptionskip}{0.2 in}
\begin{figure*}[t]
\centering
\begin{subfigure}[]{1.5 in}
\includegraphics[width=1.5 in,height=1.5 in]{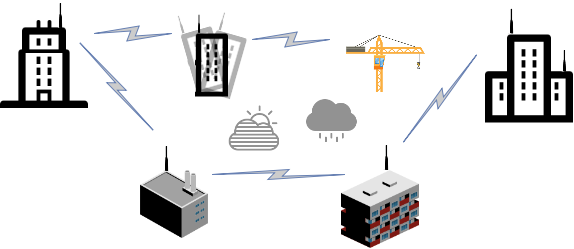}
\caption{Topology}
\label{BH1}
\end{subfigure}
~
\begin{subfigure}[]{2.2 in}
\includegraphics[width=2.3 in,height=1.5 in]{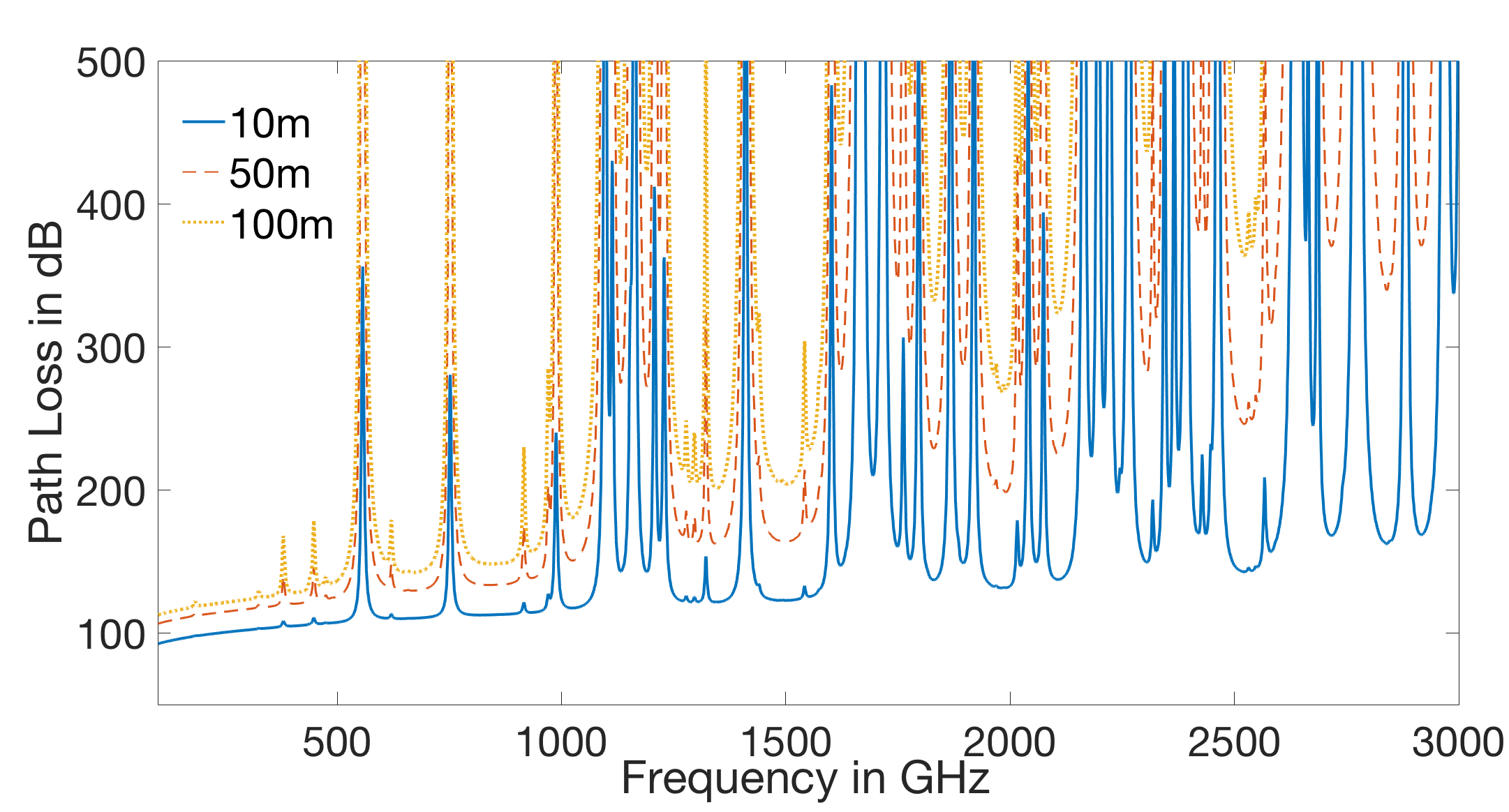}
\caption{Path Loss at RH = $50\%$}
\label{BH2}
\end{subfigure}
~
\begin{subfigure}[]{2 in}
\includegraphics[width=2.3 in,height=1.5 in]{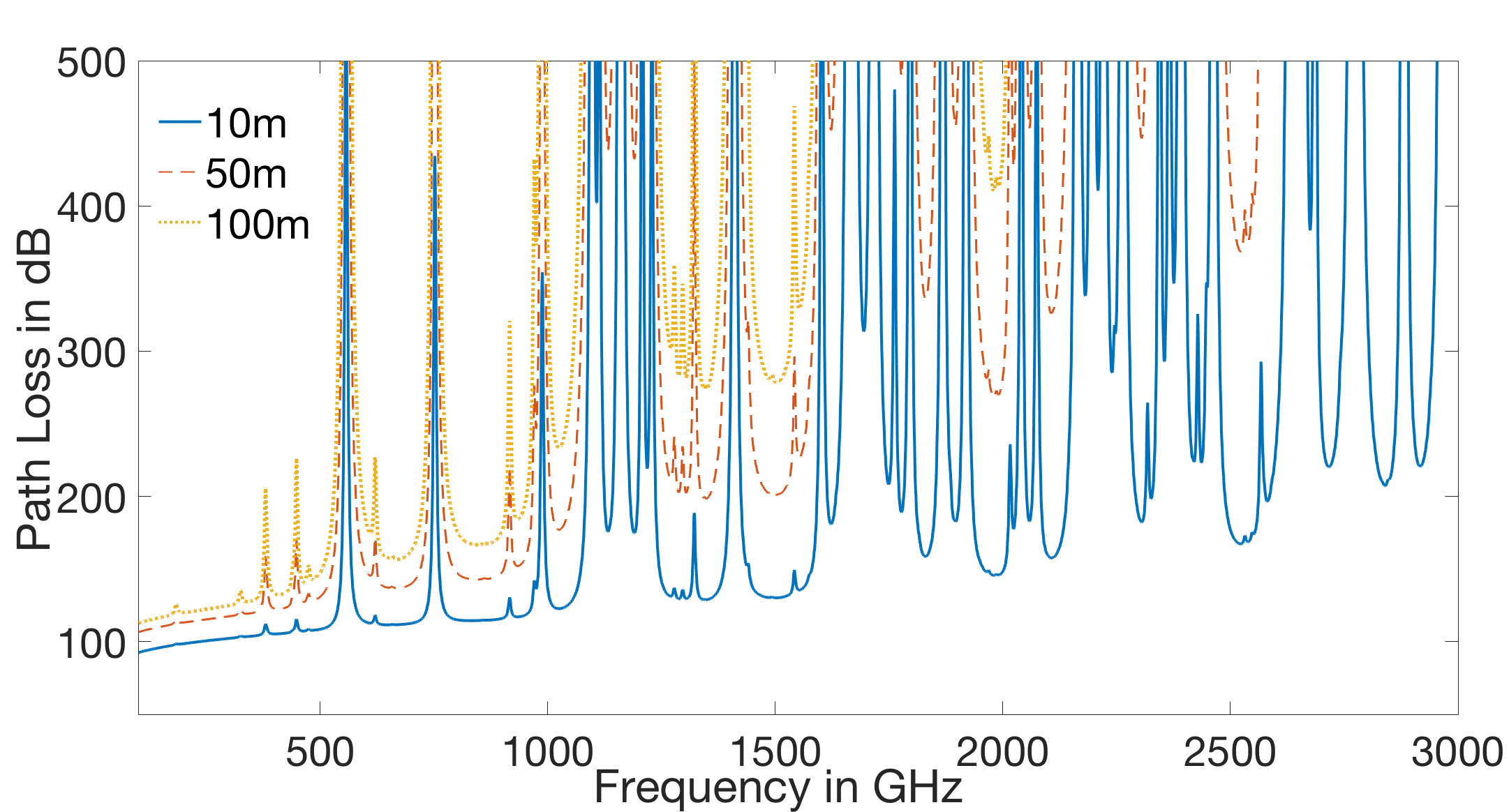}
\caption{Path Loss at RH = $100\%$}
\label{BH3}
\end{subfigure}

\caption{Wireless backhaul communication pathway and challenges } 
\label{BA}
\end{figure*}


\section{Outdoor Communication in THz Spectrum} \label{Out}

\setlength\belowcaptionskip{0 in}
\setlength\abovecaptionskip{0in}
\begin{table}[t]  
\caption{THz outdoor application classification based on AP and UE mobility types (S: Static and M: Mobile)}
\centering
\begin{tabular}{ |p{0.5cm}|p{0.6cm}|p{0.6cm}|p{4.8cm}|} 
\hline
\multirow{2}{*}{\textbf{Type}} & \multicolumn{2}{c|}{\textbf{Mobility Type}}   & \multirow{2}{*}{\textbf{Scenario Example}} \\ \cline{2-3} 
					& \textbf{AP} & \textbf{UE}  & \\ \hline 

1 & S & S & Long Range, Wireless Backhaul, Quasi-Mobile, Outdoor Displays \\ \hline
2 & S & M & Medium Range, Kiosks, Smart Bus Stops, Intelligent Transport Systems (ITS), Nomadic use\\ \hline
3 & M & S & Small range, Drone Backhaul, Aerial Base Station (ABS)\\ \hline
4 & M & M & MANET, Device to Device (D2D) communication, ABS \\ \hline					

 \end{tabular}
\label{Tab2}
\vspace{-6mm}
\end{table}

In most scenarios THz outdoor deployment will face multiple issues, such as environmental uncertainties, impenetrable obstacles, reduced user coverage, and high-scale movement and mobility \cite{OurGC, PlasticMat, RainSnow}. For reliable ultra-high outdoor communication, we require some basic properties: high bandwidth, low signal-to-noise ratio (SNR), efficient deployment, easy to install and maintain, and secure communication. THz can provide all of these properties under certain conditions and scenarios. THz deployment is justified in scenarios where benefits supersede the embedded costs. Nevertheless, in scenarios with considerable environmental uncertainty, low user demand, high mobility, and random blockage, THz use is not justified.

To address user mobility, THz devices will require fast beam alignment and device orientation detection. This might be difficult to attain for users with random movement patterns. THz systems will have to monitor small-scale \cite{SmallScale} mobility not only in the $x$, $y$, and $z$ planes but also the yaw, pitch, and roll movements of the device \cite{OurGC}. Given that mobility is one of the sensitive factors for THz deployment, we categorize the use cases for outdoor THz by the relative mobility of APs and UEs, as shown in Table \ref{Tab2}. We identify use cases where the devices are either static or mobile or a combination of both. As the relative mobility between the AP and UE increase, the communication range decreases. Each use case has its critical parameters that can be modeled. We can use adaptive systems to change these parameters on-demand basis or identify the optimal bounds for the parameters to model their topology. For example, THz can cater to \textit{quasi-mobile} applications, i.e., offloading static or less mobile users operating in the lower bands to THz bands to make room for these highly mobile users \cite{OurITS}. This form of offloading needs to be smart and triggered by: (a) the high demanding user coverage, and (b) available THz spectrum.



\section{Use Case Analysis} \label{Ana}

In this section, we conduct some preliminary analyze for a subset of the use cases introduced in Table \ref{Tab2}.

\subsection{Wireless Backhaul} \label{WBack}

For 5G or even 6G, a reliable backhaul is required to support a high-speed low-latency fronthaul. Most of the backhaul is right now supported fiber optic cables. Fiber has many challenges, such as costly infrastructure, a need for permission, and the investment is a sunk cost. On the other hand, wireless backhaul can be deployed very quickly, but the connection is uncertain dependent on environmental conditions. Free space optics (FSO) is already being considered as a means for wireless backhaul since the devices are cheap and easy to deploy. As explained in \cite{OurITS}, there are multiple challenges associated with visible light and infrared bands, and it might be beneficial to consider THz for a reliable wireless backhaul. 

The topological placement of wireless backhaul is shown in Fig. \ref{BH1}.  Environmental factors such as rain, fog, physical obstacles, and building swinging due to wind can be critical for reliable backhaul. One solution is to have a mesh of transmitter with repeaters, that allows connections to have alternative paths during environmental uncertainties \cite{SelfBack}. THz can either be used to replace fiber cables of used along with fiber cables. Replacing fibers with wireless backhaul requires direct LOS communication with maximum separation as possible. However, the separation is highly sensitive to humidity, operating frequency, bandwidth, and antenna beamwidth. 

In Fig. \ref{BH2} and \ref{BH3} we compare the total path loss (spreading loss and absorption loss) for different relative humidity (RH) and separation \cite{ChannelMod}. The spreading loss quadratically increases with the frequency, thus constraining the operation of long-range backhaul to lower frequencies of the THz band. The absorption loss spikes up multiple folds depending on the operating frequency and the humidity concentration. As the Euclidian separation and the relative humidity increases the number of frequency windows (a window is frequency gap between the spikes) increases, which results in a decreased contiguous bandwidth. The achievable data rate for the backhaul system will depend on the operating frequency, antenna gain, and location of deployment, which will directly decide the number of repeaters required between the source and destination. For example, compared to FSO based backhauls, THz can still operate in wider antenna beamwidths, thus eliminating the need for realignment when tall buildings sway due to wind. However, using wider beamwidths will reduce the throughput and increase the number of repeaters. Backhaul for smart cities should operate in frequencies below $3THz$, and can have a separation as low as $10-50m$. 


\setlength\belowcaptionskip{-0.2 in}
\setlength{\abovecaptionskip}{0.1 in}
\begin{figure}[h]
\centering
\includegraphics[width=3.5 in,height=1.3 in]{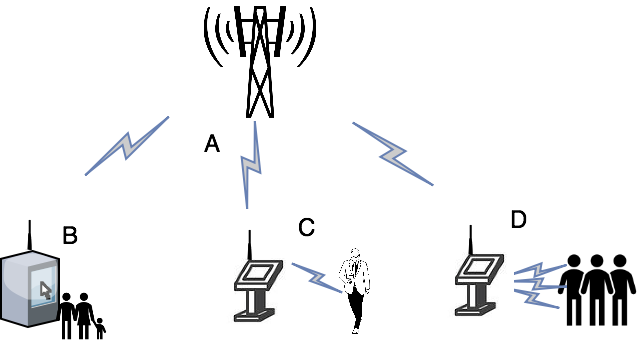}
\caption{Kiosk topology classification based on types of links.}
\label{KioArc}
\end{figure}

\setlength\belowcaptionskip{-0.2 in}
\setlength{\abovecaptionskip}{0.2 in}
\begin{figure*}[h]
\centering
\begin{subfigure}[]{2 in}
\includegraphics[width=2.2 in,height=1.8 in]{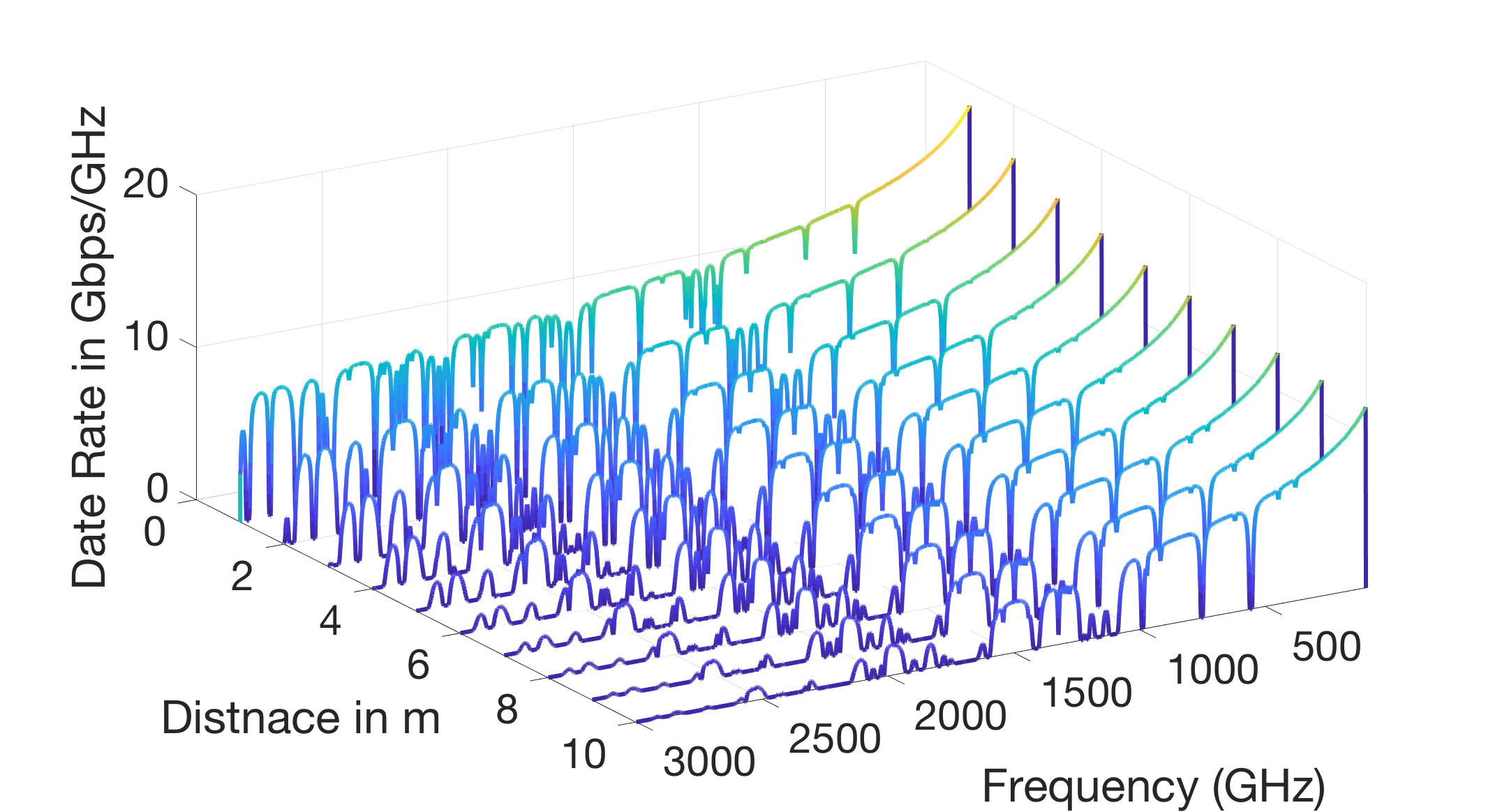}
\caption{RH = $20\%$}
\label{TA}
\end{subfigure}
~
\begin{subfigure}[]{2 in}
\includegraphics[width=2.2 in,height=1.8 in]{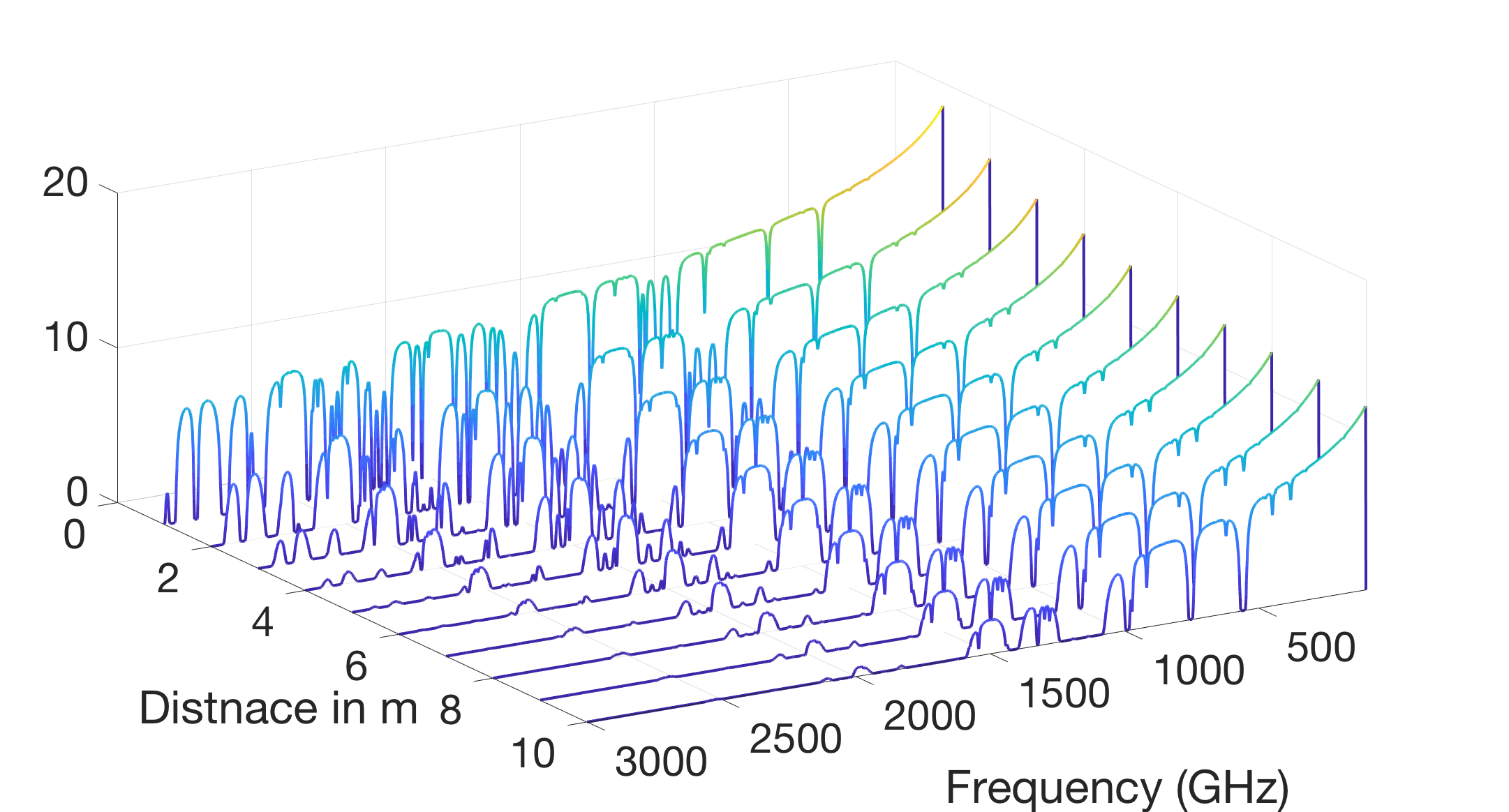}
\caption{RH = $50\%$}
\label{TB}
\end{subfigure}
~
\begin{subfigure}[]{2 in}
\includegraphics[width=2.2 in,height=1.8 in]{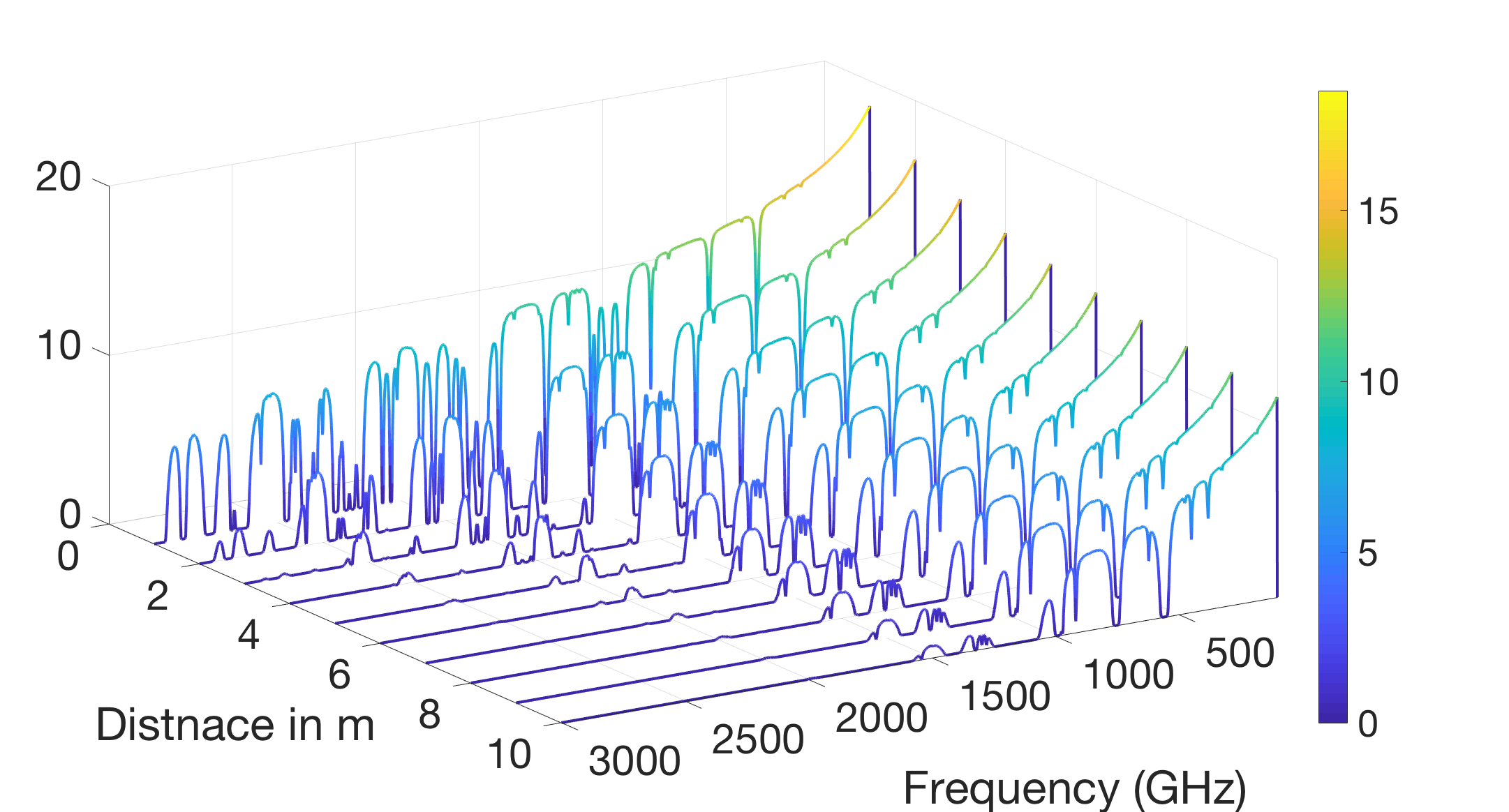}
\caption{RH = $100\%$}
\label{TC}
\end{subfigure}
\caption{Achievable data rate for varying relative humidity (RH) and distance, with fixed antenna beamwidth of $10 \degree$ } 
\label{Kio1}
\end{figure*}

\setlength\belowcaptionskip{-0.2 in}
\setlength{\abovecaptionskip}{0.2 in}
\begin{figure*}[t]
\centering
\begin{subfigure}[]{2 in}
\includegraphics[width=2.1 in,height=1.5 in]{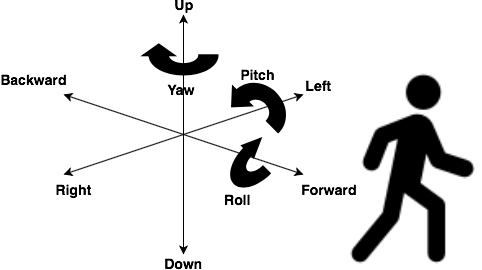}
\caption{Human mobility degree of freedom}
\label{MobiDOF}
\end{subfigure}
~
\begin{subfigure}[]{2 in}
\includegraphics[width=2.1 in,height=1.6 in]{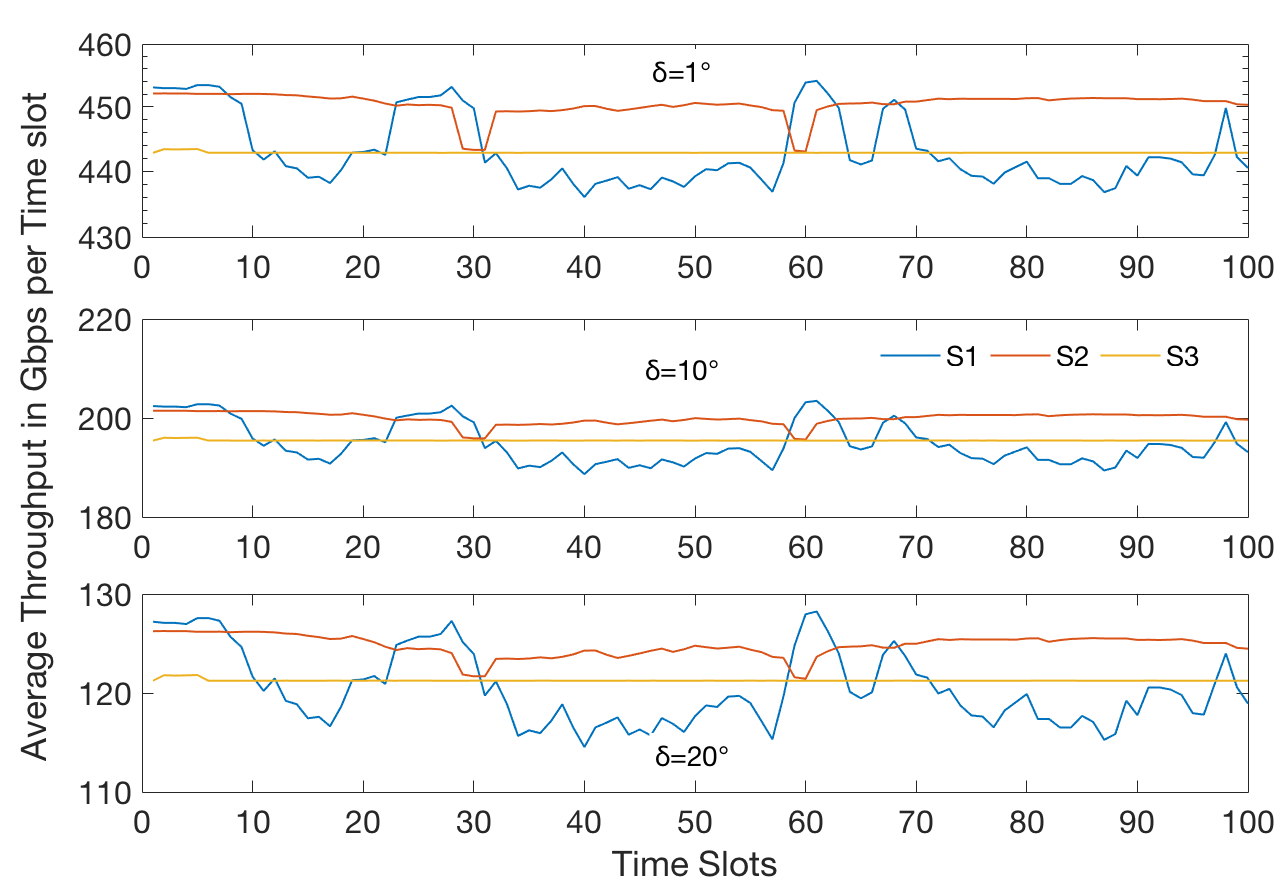}
\caption{Link C- Single User Setting}
\label{SU}
\end{subfigure}
~
\begin{subfigure}[]{2 in}
\includegraphics[width=2.1 in,height=1.5 in]{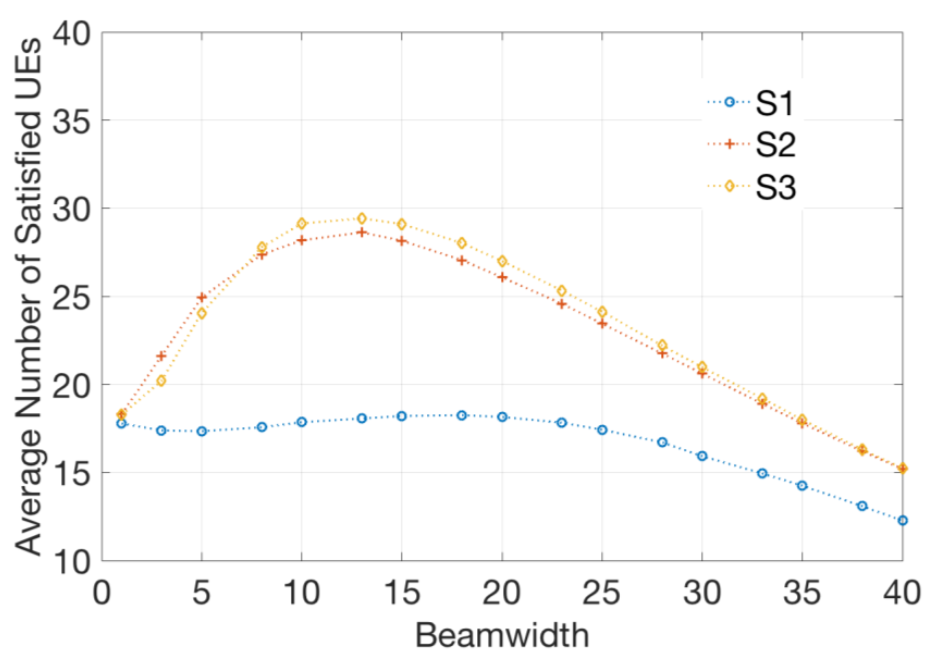}
\caption{Link D- Multi User Setting}
\label{MU}
\end{subfigure}

\caption{Parameter modeling for mobile users in a High-speed Kiosk with Link type C and D} 
\label{Kio3}
\end{figure*}

\subsection{High Speed Kiosk} \label{KioSec}

With THz, it might be challenging to cater to high-speed mobile users. However, users that require high-speed throughput can move to a nearby static kiosk \cite{KioskRef}. Although these ultra-high kiosks can be used for short-range users, the communication links can still be susceptible to small-scale mobility, such as the orientation of the device or the movement of the human body part on which the device has been mounted \cite{OurGC, SmallScale}. To deal with this challenge, we propose a kiosk-topology, as shown in Fig \ref{KioArc}, and classify the application types based on the link types.

Link A is a direct backhaul channel to the kiosk stands, which is similar to the one analyzed in section \ref{WBack}, but for a shorter range ($< 10m$). In Fig \ref{Kio1}, we analyze the minimum backhaul separation required to achieve higher throughput.  The analysis shows that we can achieve as high as $20 Gbps/GHz$ at the lower side of the THz spectrum, but the data rate decreases as we move higher in the spectrum. Moreover, the effect only becomes acute with increased relative humidity. Alternatively, we can improve the antenna gain with a narrower antenna beamwidth of $<10\degree$.

\setlength\belowcaptionskip{-0.2 in}
\setlength{\abovecaptionskip}{0.2 in}
\begin{figure*}[t]
\centering
\begin{subfigure}[]{2 in}
\includegraphics[width=2.2 in,height=1.5 in]{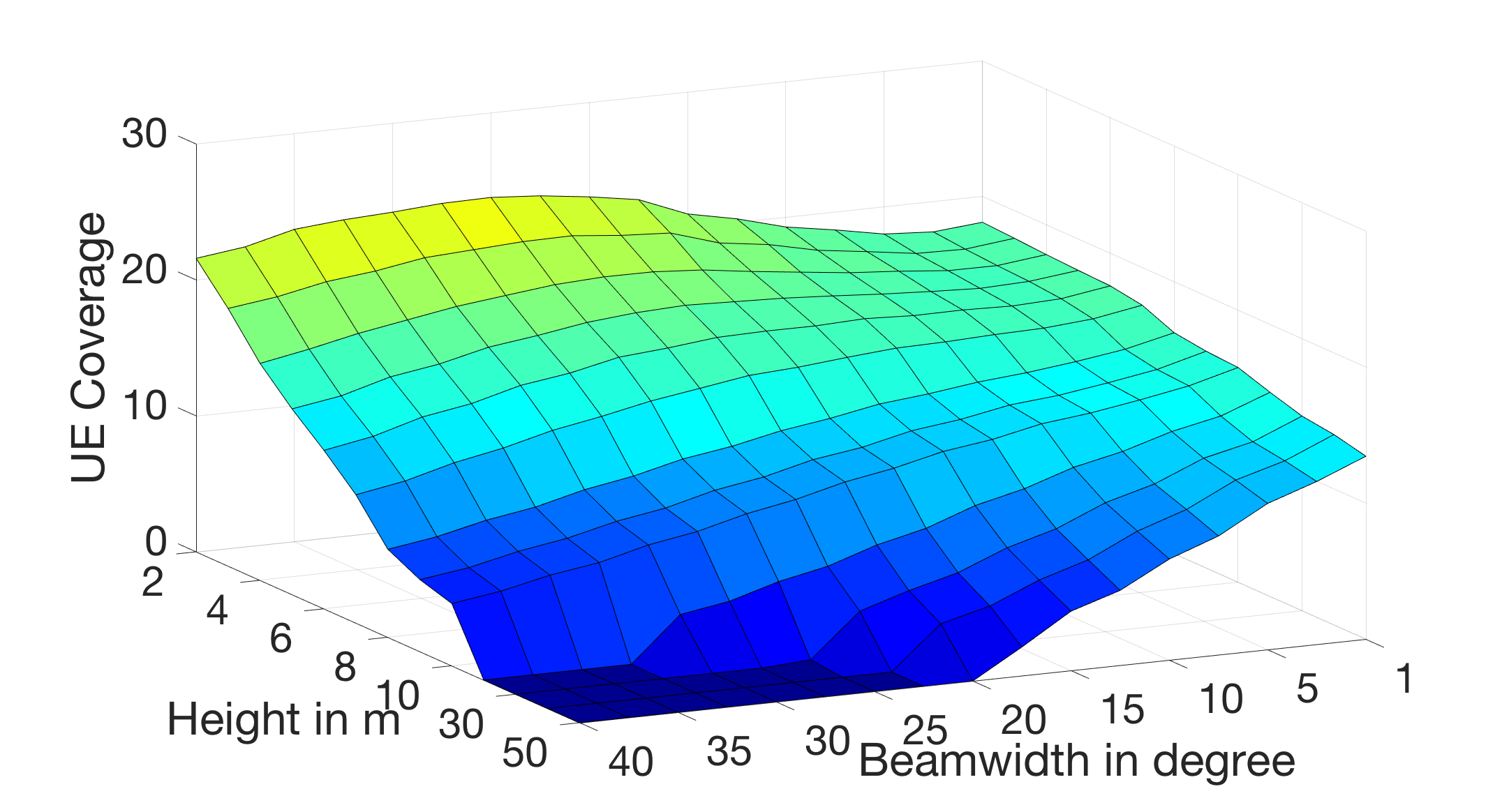}
\caption{S1 Type Users}
\label{ABS1}
\end{subfigure}
~
\begin{subfigure}[]{2.2 in}
\includegraphics[width=2.2 in,height=1.5 in]{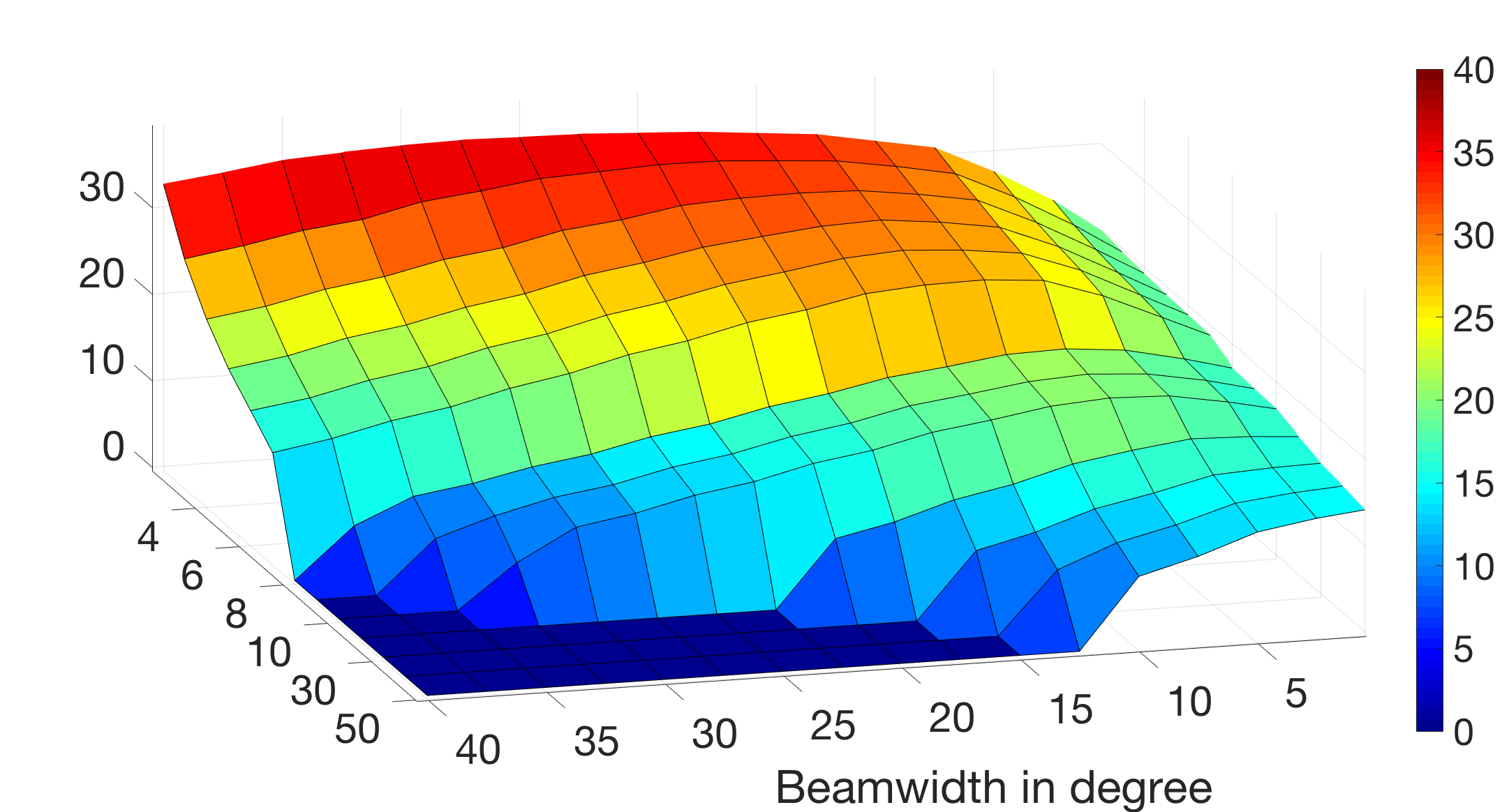}
\caption{S2 Type Users}
\label{ABS2}
\end{subfigure}
~
\begin{subfigure}[]{2.2 in}
\includegraphics[width=2.1 in,height=1.3 in]{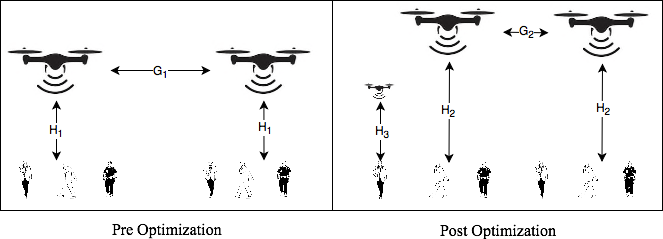}
\caption{ABS optimization example}
\label{ABS3}
\end{subfigure}

\caption{Parameter modeling for aerial base stations} 
\label{ABSa}
\end{figure*}


The kiosk stands might be of three types: (a) Link B- informative with high definition video or an interactive framework for users, (b) Link C- high-speed throughput for a single user, or (c) Link D- high-speed throughput for multiple-users. Link B is a visual interaction with the user and does not involve any communication link. To avoid mobility induced outages, for links C \& D, we have to monitor small-scale user movements in all six-degrees-of-freedom (DoF), as shown in Fig. \ref{MobiDOF}. Since small orientational change in each axis can cause significant outages, we classify our analysis into three mobility types, namely: (a) S1: High Mobility (e.g., high intense), (b) S2: Constrained Mobility (e.g., slow walking), and (c) S3: Low Mobility (e.g., sitting or standing) \cite{RobotExercise}. We assume an oscillation in the yaw, pitch and roll axises of $13\degree-15\degree$ for S1, $5\degree-3\degree$ for S2, and $3\degree-1\degree$ for S3. In the simulation results shown in Fig. \ref{SU} \& \ref{MU} we assume that the frequency and the available bandwidth is selected adaptively based on the humidity and distance.

For link C connection, a Kiosk has to maintain a connection with a single user, and the average throughput varies based on the mobility type, as shown in Fig. \ref{SU}. The variations in the average through are due to the constant beam misalignments and realignment required for different service types. Although decreasing the antenna beamwidth $\delta$ improves the throughput in link C, it is not the same for link D connection with multiple users. In link D connection, for users with the same antenna beamwidth and demanding a minimum throughput of $10 Gbps$, we observe optimal beamwidth values that maximize user coverage, as shown in Fig. \ref{MU}. The varying oscillations for mobility type S1 to S3 results in different optimal beamwidth values. To improve link connectivity and decrease outage time, we need fast beam alignment strategies. Alternatively, we could use NLOS communication by strategically placing reflective surfaces to improve coverage, but this becomes a burdensome deployment approach.


\subsection{Aerial Base Station}

The aerial base station (ABS) \cite{aerialBS} is a novel concept where drones are used as base stations to cater to users with wireless services. ABS is useful in scenarios of disaster management, public safety, and on-demand congestion control. It can also provide flexibility for wireless backhaul by obtaining an alternative path during bad weather conditions. At a considerable altitude, with low temperature and very low humidity, the THz spectrum can effectively be used by drones to form flying ad-hoc networks (FANET) \cite{FANET}. THz-drones are already being used for spectroscopy purposes of detecting atmospheric molecules and can be easily extended for communications purposes. ABS will be very sensitive to the altitude and relative mobility, which can either case 3 or 4, as shown in Table \ref{Tab2}. For mobile users, we need perfect and fast beam alignment to prevent outages. In Fig. \ref{ABS1} and \ref{ABS2}, we analyze the optimal height of the drones for user mobility types S1 and S2, introduced in section \ref{KioSec}. For a multi-user setting, our analysis shows that there exist optimal height and antenna beamwidth values where a THz-drone can achieve high throughput even with a limited bandwidth of $10GHz$. In the future, we might need self-optimizing THz-drones that adjust the height and the inter-drone gap based on the user demand, as shown in Fig. \ref{ABS3}. 



\section{Conclusion} \label{Conclu}

In this paper, we propose different settings for outdoor THz deployment. There exist multiple use-cases where reliable communication is possible; however, it requires strategic deployment and opportunistic parameter modeling. We show that future THz devices will require systems that can opportunistically change parameters, such as antenna beamwidth, antenna alignment, antenna height, and operating frequency. To understand and optimize the reliability of the use-cases, we model the outdoor-based applications using a relative mobility model between the APs and the UEs. It is evident that THz alone will not be able to solve all the issues in outdoor wireless communication and will require the support of lower frequency bands. For example, in a quasi-mobile application, we might need a queuing model that can decide on how many static applications should be offloaded to THz spectrum to make room for one high-mobile user in the lower bands. In the future, we will like to design strategies based on specific environmental factors, user movement patterns, and possibly use advanced learning algorithms to improve reliability. 


\end{document}